\documentclass[%
 aip,
 amsmath,amssymb,
preprint,%
]{revtex4-1}

\usepackage{graphicx}
\usepackage{dcolumn}
\usepackage{bm}
\usepackage[mathlines]{lineno}

\usepackage[utf8]{inputenc}
\usepackage[T1]{fontenc}
\usepackage{mathptmx}
\usepackage{etoolbox}
\usepackage[caption=false]{subfig}
\usepackage{float}
\usepackage{subfig}
 \DeclareMathOperator{\diag}{diag}
\makeatletter
\def\@email#1#2{%
 \endgroup
 \patchcmd{\titleblock@produce}
  {\frontmatter@RRAPformat}
  {\frontmatter@RRAPformat{\produce@RRAP{*#1\href{mailto:#2}{#2}}}\frontmatter@RRAPformat}
  {}{}
}%
\makeatother
\begin{document}

\preprint{AIP/123-QED}

\title{Onset of convective instability in an inclined porous medium\\}
\author{Emmanuel E. Luther}
\email{luthere@coventry.ac.uk.}
 \affiliation{ 
Fluid and Complex Systems Research Centre, Coventry University, United Kingdom
}%
\author{Michael C. Dallaston}
\email{michael.dallaston@qut.edu.au}
\affiliation{%
School of Mathematical Sciences, Queensland University of Technology, Brisbane
}%
\author{Seyed M. Shariatipour}%
\email{ab6995@coventry.ac.uk}

\author{Ran Holtzman}%
\email{ad2472@coventry.ac.uk}
\affiliation{ 
Fluid and Complex Systems Research Centre, Coventry University, United Kingdom
}%

\date{\today}

\begin{abstract}
The diffusion of a solute from a concentrated source into a horizontal, stationary, fluid-saturated porous medium can lead to a convective motion when a gravitationally unstable density stratification evolves. In an inclined porous medium, the convective flow becomes intricate as it originates from a combination of diffusion and lateral flow, which is dominant near the source of the solute. Here, we investigate the role of inclination on the onset of convective instability by linear stability analyses of Darcy's law and mass conservation for the flow and the concentration field. We find that the onset time increases with the angle of inclination ($\theta$)  until it reaches a cut-off angle beyond which the system remains stable. The cut-off angle increases with the Rayleigh number, $Ra$. The evolving wavenumber at the onset exhibits a lateral velocity that depends non-monotonically on $\theta$  and linearly on $Ra$. Instabilities are observed in gravitationally stable configurations ($\theta \geq 90^{\circ}$) solely due to the non-uniform base flow generating a velocity shear commonly associated with Kelvin-Helmholtz instability. These results quantify the role of medium tilt on convective instabilities, which is of great importance to geological CO$_2$ sequestration.

\end{abstract}

\maketitle

%

\section{\label{sec:level1}introduction} 

In a fluid-saturated porous medium where gravity is dominant, an unstable stratification of a dense solution overlaying a less dense one may generate a convective motion. This phenomenon, first observed by Horton and Rogers \cite{HortonandRogers1945} and Lapwood \cite{Lapwood1948}, is important in applications including CO$_2$ geological sequestration\cite{HassanzadehKeithPooladi-Darvish2005}, evaporative salt convection \cite{WoodingTyler1997}, contaminant transport \cite{BearCheng2010}, and building insulation \cite{StoreslettenPop1996,BanksandZaturska1991}. Convective instability is considered steady when the buoyancy source creates a constant density stratification, and transient when the density profile evolves in time. A detailed review of porous medium convection discussing the steady and transient cases in horizontal and inclined configurations is outlined in Nield and Bejan \cite{NieldBejan2013} and Hewitt \cite{Hewitt2020}. While convective instability has been extensively studied in horizontal systems (for both the steady\cite{HortonandRogers1945,Lapwood1948} and transient cases\cite{HassanzadehKeithPooladi-Darvish2005,HassanzadehKeithPooladi-Darvish2007,TiltonDanielRiaz2013,RiazHesseTchelepiOrr2006,Slim2014,RaadHassanzadeh2015,XuChenZhang2006,Lutheretal.2021}), only few studies addressed inclined porous media, especially in the transient\cite{JavaheriAbediHassanzadeh2008,JavaheriAbediHassanzadeh2009} case which occurs in many applications.

In horizontal systems, the stability of the time-dependent diffusive base state has been analyzed using the initial value problem (IVP) approach and the quasi-steady state approximation (QSSA) \cite{TanHomsy1986}. IVP determines the stability character of the porous system by numerically evaluating the growth rates of concentration ($\sigma_c$) and velocity perturbations ($\sigma_v$) without assuming the base state to be instantaneously steady. This approach is sensitive to initial conditions and depends on how the growth rate is measured, introducing subjectivities \cite{TiltonDanielRiaz2013,TanHomsy1986}.  

On the other hand, QSSA predicts the onset time of instability and the unstable disturbance modes, by examining the growth rate ($\sigma$) in time ($t$) of various modes of small disturbances ($\omega$) in space ($x$) for both concentration and velocity, over a transient base state assumed to be instantaneously steady \cite{TiltonDanielRiaz2013}. The accuracy of the assumption that the diffusive base state can be held instantaneously constant in QSSA improves with time as diffusion becomes slower. The upper bound of the validity period of this approximation is when diffusion breaks down, becoming invalid in the non-linear regime, and is estimated to be when $\sigma t\sim1$, representing when the initial perturbations start becoming large \cite{TrevelyanAlmarchaDeWit2013}. The disturbance growth rate when QSSA is implemented in self-similar coordinates, a moving reference frame, compare reasonably well at short time with dominant mode, a method valid during early time, and with IVP for all regimes \cite{RiazHesseTchelepiOrr2006}. This agreement in the onset time is because the growth rate in self-similar QSSA is compared with the growth rate of concentration perturbation ($\sigma_c$) in IVP. However, when the growth rate of velocity perturbation ($\sigma_v$) is considered, the onset predictions in IVP approximate onset time in normal QSSA, the stationary cartesian coordinate  system. Interestingly, both $\sigma_c$ and $\sigma_v$ converge to the value of $\sigma$ in normal QSSA at later time \cite{TiltonDanielRiaz2013}. The scaling of the linear onset time $t_o =\alpha_0 D^{-1} Ra^{-2} H^{2}$ for the scaling prefactor ($\alpha_0$) values of $36\leq \alpha_0 \leq 143$ and Rayleigh number ($Ra$) range of $100 \leq Ra \leq 8000$ is similar for IVP ($\sigma_c$, $\sigma_v$), normal ($\sigma$) and self-similar QSSA, where $H$ is the thickness and $D$ is the effective diffusion coefficient of the porous system \cite{TiltonDanielRiaz2013}. The Rayleigh number ($Ra$) describes the relative importance of convection to diffusion (we provide the precise definition of $Ra$ in the next section). Furthermore, the onset time and critical wavenumber from QSSA compare well with the modes in numerical simulations \cite{Slim2014}.

The literature on the stability of a time-dependent density profile is preceded by a great deal of research on the stability of a steady, linear temperature profile when a horizontal fluid saturated porous medium is heated from below and cooled at the top simultaneously \cite{HortonandRogers1945,Lapwood1948}. For this profile in an isotropic homogenous system with impermeable and isothermal boundaries, the critical Rayleigh number above which a steady conduction state becomes unstable is determined analytically to be $Ra=4\pi^2$. The critical conditions for various modifications of the boundaries  are outlined by Nield and Bejan \cite{NieldBejan2013}. 

\begin{figure*}
\includegraphics[
  width=18cm,
  height=6cm,
  keepaspectratio,
]{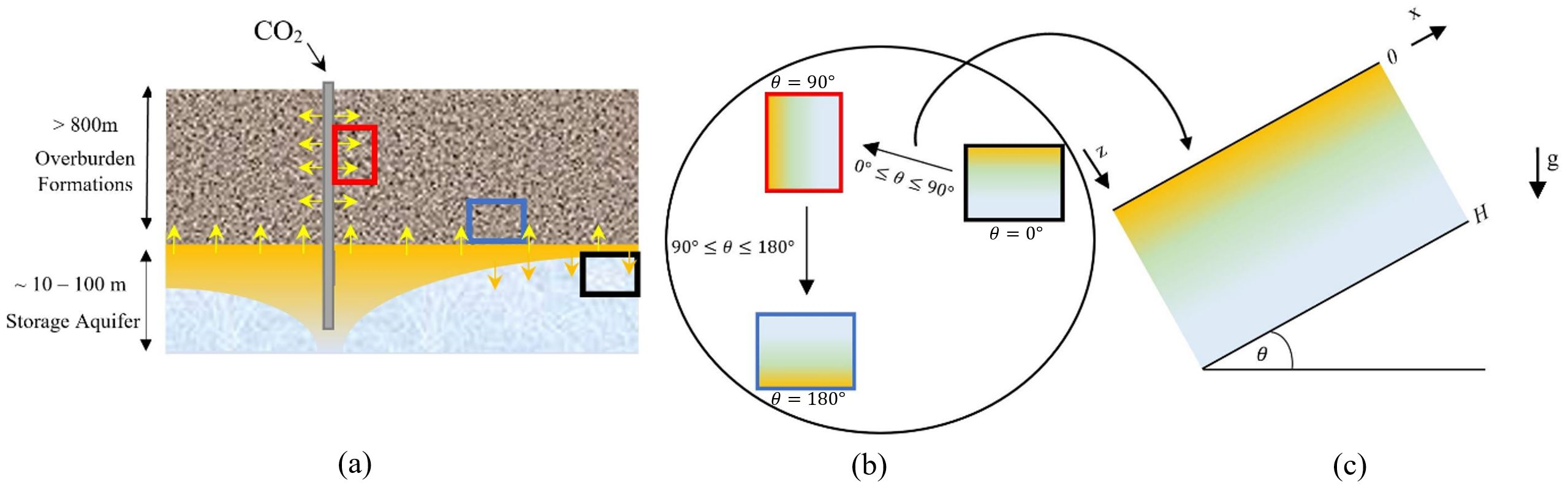}
\caption{\label{fig:inclined1}Schematics of the model under consideration. (a) The diffusion of dissolved CO$_2$ from the saturated pore fluid in contact with the buoyant plume into less saturated zones in the storage aquifer, from the wellbore into the overburden formations, and from the storage aquifer into the overburden formations. (b) The inclinations describing the 3 diffusion scenarios, with the intermediate angles accounting for tilts in the storage aquifer and deviations of the injection well. (c) An idealized model inclined at an angle $\theta$, which is within the range of inclinations considered. The colors indicate CO$_2$ declining from yellow to blue, while the yellow arrows in (a) depict the diffusive transfer of dissolved CO$_2$. The overburden formation is separated from the storage aquifer and the wellbore by a sharp diffusive interface.}
\end{figure*}

In an inclined porous layer with impermeable boundaries, the stability of a constant temperature profile is associated with various flow structures depending on $Ra$, the angle of inclination $\theta$ and the dimensions of the domain whether two - (2D) or three - dimensional (3D). The unicellular basic flow observed in experiments comprises an upward and downward fluid motion along the bottom (heated) and the top (cooled) plate respectively \cite{BoriesCombarnous1973}, which becomes unidirectional if the layer was laterally infinite \cite{WenChini2018}. The unicellular flow has no convective pattern specified by zero transverse ($\omega_x= 0$) and longitudinal ($\omega_y= 0$) wavenumber in a 3D domain \cite{WenChini2018}. For $Ra\cos\theta>4\pi^2$, experimental results suggest that the unicellular or unidirectional mode with a steady temperature profile transforms to a polyhedric ($\omega_x\neq 0$ ,$\omega_y\neq 0$) or transverse roll ($\omega_x\neq 0$,$\omega_y = 0$)  for $\theta \leq 15^{\circ}$ but to a helicoidal mode ($\omega_x =  0$,$\omega_y\neq 0$) for larger inclinations \cite{BoriesCombarnous1973}. Theoretical analysis predicts that the transition from polyhedric or transverse rolls to helicoidal mode occurs at $\theta \approx 31.8^{\circ}$ when the condition on $Ra\cos\theta$ is met \cite{CaltagironeBories1985}. A theoretical study in a 2D domain where the longitudinal wavenumber ($\omega_y$) is non-existent reveals that a precise condition for the occurrence of instability is when $\theta \leq 31.8^{\circ}$ for large $Ra$ and $\theta \leq 31.49^{\circ}$ in general \cite{WenChini2018}, supporting the previous findings that a porous vertical slab ($\theta = 90^{\circ}$) is always stable \cite{Gill1969}. Above this critical cut-off value of $31.49^{\circ}$, the system is stable for any $Ra$. 

The result in inclined porous media depends on the boundary conditions of the model such as permeable \cite{Barletta2015,BarlettaCelli2018,CelliBarletta2019}, uniform flux\cite{BarlettaStoresletten2011,ReesBarletta2011}, nonuniform linear temperature distributions \cite{BarlettaRees2012}, and time periodic temperature\cite{WuWang2017}; it depends on the model flow equations such as Forchheimer \cite{ReesPostelnicuStoresletten2006} and Brinkman equations \cite{FalsaperlaGiacobbeMulone2019}, extending Darcy’s law, and on other relaxed assumptions including viscous dissipation \cite{NieldBarlettaCelli2011}, anisotropy \cite{StoreslettenRees2019}, heterogeneity\cite{McKibbin2014,NieldKuznetsovBarlettaCelli2016}, double diffusive convection \cite{KumarNarayanaSahu2020,DubeyMurthy2020}, and local thermal non-equilibrium \cite{BarlettaRees2015,NieldKuznetsovBarlettaCelli2016}. A summary on convective instability in inclined systems is reported by Nield\cite{Nield2011} and Nield and Bejan\cite{NieldBejan2013}.

The earlier work on the stability of a time-dependent profile in an inclined configuration analyzed the case where the base state is stationary\cite{JavaheriAbediHassanzadeh2008,JavaheriAbediHassanzadeh2009}. However,
there are 3 outstanding questions in the literature on the analysis of the stability of a transient base state which we address in this paper: (i) how does the angle of inclination affect the onset time and perturbation dynamics? (ii) what is the cut-off angle beyond which the system remains stable? (iii) what is the implication when $\theta \geq 90 ^{\circ}$ in which the system is gravitationally stable? .

 Therefore, we investigate the stability of a transient diffusive concentration profile in an inclined porous layer in the context of CO$_2$ geological storage due to its relevance and important practical implications. Inclination introduces a near boundary buoyancy-induced flow which may influence the dynamics during the formation of instabilities, and it can affect the driving force of the convection. We investigate this problem in 2D to obtain insights into the evolution of perturbations, to reduce computational demands and to augment the current understanding on porous media convective instability in 2D horizontal systems. For the first time, we observe a $Ra$ – dependent cut-off angle $\theta$, beyond which the system does not become unstable. In addition, the inclinations delay the onset time and the growing instabilities are in lateral motion. In contrast to the steady base state, any angle $\theta \leq 180^{\circ}$ will be unstable for sufficiently large $Ra$.  

\section{\label{sec:level1}model description and governing equations}

We consider an idealized porous layer, inclined at an angle $\theta$ to the horizontal, assumed to be laterally infinite, homogeneous, isotropic and saturated with liquid-phase (Fig.~\ref{fig:inclined1}). The layer comprises of an impermeable top boundary with a diffusing CO$_2$ source  and a bottom boundary that is impermeable to both CO$_2$ and the pore fluid. The study is restricted to the parameter range $10^2\leq Ra \leq 10^3$ typical for storage aquifers \cite{HassanzadehKeithPooladi-Darvish2007}  and applicable to the overburden formations for some realistic aquifer–to–overburden parameter ratios of permeability – $k_a / k_o \sim 10^3$, diffusivity – $ D_a/ D_o\sim10$, porosity – $\phi_a/ \phi_o\sim 10$, and thickness –  $H_a/ H_o \sim 10^{-1}$, where the subscripts $a$ and $o$ denote aquifer and overburden respectively \cite{DejamHassanzadeh2018}. The properties of the fluid saturating the aquifer and the overburden formation such as the viscosity ($\mu$) and density ($\rho$) are assumed to be identical. 

Geological formations are usually structurally heterogenous over multiples spatial scales. Therefore, the following assumptions are made to reduce the complexity of the problem to focus on the formation of instabilities in various homogeneous configurations due to a diffusive transfer from a concentrated source: we assume a sharp diffusive interface between the overburden formations  and (i) the storage aquifer and (ii) the wellbore, with the interface diffusive transfer controlled by $D_o$, the overburden diffusion coefficient. It is common for dissolved CO$_2$ to leak by diffusion from storage reservoirs \cite{KroossBuschAllesAmann-Hildenbrand2003,FleuryBerneBachaud2009,BuschAmann-HildenbrandBertier2010} and fractured wellbores \cite{DuguidRadonjicScherer2011,UmKabilanSuhFernandez2014} into overburden formations. Hence, the range of inclinations $0^{\circ} \leq \theta \leq 180^{\circ}$ are considered to model different scenarios during CO$_2$ sequestration (Fig.~\ref{fig:inclined1}) in a tilted storage aquifer \cite{,TsaiRiesingStone2013,BharathSahuFlynn2020} with deviated injection wells \cite{OkwenStewartCunningham2011,MathiesonMidgelyWright2011,NewellMartinezEichhubl2017}.

Further simplifying assumptions include the consideration of the flow of an incompressible, single phase fluid; and the neglection of geochemical reactions, background flow of groundwater, other sources, and sinks. The flow is governed by Darcy’s law, and the density, which is linearly dependent on the solute concentration, varies according to Boussinesq approximation only in the buoyancy term ($\rho g$). The fluid motion in these settings is governed by the following equations \cite{HassanzadehKeithPooladi-Darvish2005,RiazHesseTchelepiOrr2006,XuChenZhang2006} 

\begin{equation}
\frac{\partial u}{\partial x} + \frac{\partial v}{\partial z} = 0 ,
\end{equation}
\begin{equation}
u = - \frac{k}{\mu}\left(\frac{\partial p}{\partial x}+(1+\beta c)\rho_o g\sin\theta\right) ,
\end{equation}
\begin{equation}
v = - \frac{k}{\mu}\left(\frac{\partial p}{\partial z}-(1+\beta c)\rho_o g\cos\theta\right) ,
\end{equation}
\begin{equation}
\phi\frac{\partial c}{\partial t}+u\frac{\partial c}{\partial x}+v\frac{\partial c}{\partial z} = \phi D \left(\frac{\partial^2 c}{\partial x^2}+\frac{\partial^2 c}{\partial z^2}\right) ,
\end{equation}
where $c$ is the $CO_2$ dissolved concentration, $u$ and $v$ are Darcy velocity components, $p$ is the pressure, $k$ is the permeability, $\phi$ is the porosity, $\mu$ is the viscosity, $g$ is the gravity, $\rho_o$ is the original brine density, $\beta$ is the coefficient of density variation with concentration, and $D$ is the effective diffusion coefficient of $CO_2$ in brine, which accounts for the diffusivity of $CO_2$ in brine and the rock tortuosity.  
The source of the solute at $z=0$ is modeled with a constant concentration, $c_s$, the solubility concentration of $CO_2$ in brine, and the boundary at $z=H$ has no flux, $dc/dz=0$, for $t\geq0$. The system $(0 \leq z \leq H)$ is initially assumed to contain no solutes $(c(t=0)=0)$. The following dimensionless variables are introduced: $\hat{x} = x/H$ , $\hat{z}=z/H$, $\hat{c}=c/c_s$, $\hat{u}=Hu/\phi D$, $\hat{v}=Hv/\phi D$, $\hat{t}=Dt/H^2$, $\hat{p} = k(p - \rho_o g(z\cos\theta - x\sin\theta))/\phi \mu D$, so that the dimensionless governing equations becomes:
\begin{equation}
\frac{\partial \hat{u}}{\partial \hat{x}} + \frac{\partial \hat{v}}{\partial \hat{z}} = 0 ,
\end{equation}
\begin{equation}
\hat{u} = - \frac{\partial \hat{p}}{\partial \hat{x}}- Ra\sin\theta \hat{c} ,
\end{equation}
\begin{equation}
\hat{v} = - \frac{\partial \hat{p}}{\partial \hat{z}}+ Ra\cos\theta \hat{c} ,
\end{equation}
\begin{equation}
\frac{\partial \hat{c}}{\partial \hat{t}}+ \hat{u}\frac{\partial \hat{c}}{\partial \hat{x}}+\hat{v}\frac{\partial \hat{c}}{\partial \hat{z}} = \frac{\partial^2\hat{c}}{\partial \hat{x}^2}+\frac{\partial\hat{c}^2}{\partial\hat{z}^2} ,
\end{equation}
where $Ra = k \rho_o\beta c_s  g H/\phi \mu D$, the definition commonly used in porous media convection \cite{XuChenZhang2006,RaadHassanzadeh2015,Hewitt2020,Lutheretal.2021}, which differs from that in an unconfined fluid where $\phi = 1$, and the fluid flow is governed by the Navier-Stokes equations. Our characteristic volume-averaged velocity ($\phi D/H$) is obtained by multiplying the porosity $\phi$ (the fraction of the porous medium that is void) by the scale of the actual diffusive speed of the solute in the fluid within the pores, ($D/H$).

We eliminate the pressure term by subtracting the $x$-derivative of (7) from the $z$-derivative of (6) and apply mass conservation:
\begin{equation}
\frac{\partial^2\hat{v}}{\partial\hat{x}^2}+\frac{\partial^2\hat{v}}{\partial\hat{z}^2} = Ra \left( \frac{\partial^2\hat{c}}{\partial\hat{x}^2}\cos\theta+\frac{\partial^2\hat{c}}{\partial\hat{x}\partial\hat{z}}\sin\theta \right).
\end{equation}

\section{\label{sec:level1}Linear Stability Analysis}

The flow and transport variables are decomposed into base states and perturbation variables $\hat{A}(\hat{x},\hat{z},\hat{t})= \hat{A}_b (\hat{z},\hat{t})+\hat{A'}(\hat{x},\hat{z},\hat{t})$, where $\hat{A}$ represents the flow and transports variables $(\hat{p},\hat{c},\hat{u},\hat{v})$; subscript (b) and superscript ($'$) refer to the base state and perturbed variable respectively. The base state problem requires that $\displaystyle{\frac{\partial \hat{c}_b }{\partial \hat{x}} = 0}$ since the base concentration $(\hat{c}_b)$ does not vary in the $x$-direction, with $\hat{v}_b=0$ and $\hat{u}_b= - Ra\sin\theta\hat{c}_b$ due to the unbalanced gravity term in the $x$-direction Eq.(6). The horizontal base state velocity ($\hat{u}_b$) is non-uniform varying with depth and time dependent. Though the base state is not stationary, its concentration satisfies the diffusion equation $\displaystyle{\frac{\partial^2\hat{c}_b}{\partial\hat{z}^2} = \frac{\partial\hat{c}_b}{\partial\hat{t}}}$, the boundary conditions, and the initial condition, $\hat{c}=0$ in $0<\hat{z}< 1$  for $\hat{t} = 0$ as $\hat{u}_b$ does not transport a concentration gradient $\displaystyle{ \frac{\partial\hat{c}_b}{\partial\hat{x}}= 0}$. Using separation of variables, a solution for $\hat{c}_b$ may be found as $\hat{c}_b(\hat{z},\hat{t})= \frac{4}{\pi} \sum_{m=1}^{\infty}\frac{1}{(2m-1)}e^{-\left(\frac{(2m-1)\pi}{2}\right)^2\hat{t}} \sin \left(\frac{(2m-1)\pi \hat{z}}{2} \right)$. The perturbation equations can be derived by substituting the decomposition of the variables into equations (8) and (9) while assuming the magnitude of the perturbation variables is sufficiently small such that higher order perturbation terms can be neglected:

\begin{equation}
\frac{\partial^2\hat{v}\:'}{\partial\hat{x}^2 }+\frac{\partial^2\hat{v}\:'}{\partial\hat{z}^2} = Ra \left( \cos\theta \frac{\partial^2\hat{c}\:'}{\partial\hat{x}^2}+\sin\theta\frac{\partial^2\hat{c}\:'}{\partial\hat{x}\partial\hat{z}} \right),
\end{equation}

\begin{equation}
\frac{\partial\hat{c\:}'}{\partial\hat{t}} - Ra \sin\theta\hat{c}_b\frac{\partial\hat{c}\:'}{\partial\hat{x}}+\hat{v}\:'\frac{\partial\hat{c}_b}{\partial\hat{z}} = \frac{\partial^2\hat{c}\:'}{\partial\hat{x}^2}+\frac{\partial^2\hat{c}\:'}{\partial\hat{z}^2}.
\end{equation}

Instead of seeking an arbitrary solution $(\hat{v}\:',\hat{c}\:')(\hat{x},\hat{z},\hat{t}) = (v^*,c^*)(\hat{z},\hat{t})e^{i\omega \hat{x}}$ for the time evolution of the perturbations in (10) and (11) due to the transient nature of the base state $\displaystyle{\frac{\partial\hat{c_b}}{\partial \hat{z}}(\hat{z},\hat{t})}$, we employ QSSA by assuming the separation of timescales namely that the solute diffusion is substantially slower than the growth or decay of small perturbations. To apply QSSA, the evolving perturbations can be expressed as $(\hat{v}\:',\hat{c}\:')(\hat{x},\hat{z},\hat{t}) = (v^*,c^*)(\hat{z},\hat{t_s})e^{i\omega \hat{x}+\sigma(\hat{t}_s)\hat{t}}$ assuming the base state to be instantaneously stationary $\displaystyle{\frac{\partial\hat{c}_b}{\partial \hat{t}}(\hat{z},\hat{t}_s) = 0}$ at a given time $\hat{t}_s$ since the growth of the perturbation is faster than the evolution of the base state. By substituting the QSSA expression for the perturbations into (10) and (11) with the associated boundary conditions, we can derive the ordinary differential equations 
$\left(\displaystyle{\frac{\partial^2}{\partial \hat{z}\:^2}} - \omega^2 \right)v^* = Ra \left(-\omega^2c^*\cos\theta + i\omega \displaystyle{\frac{\partial c^*}{\partial\hat{z}}}\sin\theta\right)$ and $\sigma c^* - i\omega Ra \hat{c}_b \sin\theta c^*+ v^* \displaystyle{\frac{\partial \hat{c}_b}{\partial\hat{z}}} = \left(\displaystyle{\frac{\partial^2}{\partial \hat{z}\:^2}} - \omega^2 \right)c^*$ and the boundary conditions $ v^*=c^*=0$ at $\hat{z} = 0$, $\hat{t} > 0$ and $ v^*=\displaystyle{\frac{dc^*}{d\hat{z}}} = 0$ at $\hat{z} = 0$, $\hat{t}>0$, where $\omega$ and $\sigma$ are wavenumber in the $x$-direction and growth rate of the perturbations respectively.  These equations are solved numerically in a finite difference numerical scheme after posing them in the following matrix form:

\begin{equation}
\bm{B} v^* = Ra \bm{H} c^* , 
\end{equation}

\begin{equation}
\sigma c^* - i\omega Ra \diag(\bm{\hat{c}}_{\bm{b}}) \sin\theta c^* +v^* \diag\left(\frac{\bm{\partial\hat{c}_b}}{\bm{\partial\hat{z}}}\right) = \bm{A}c^* .
\end{equation}

After some algebraic substitutions, we arrive at the eigenvalue problem:
\begin{equation}
\sigma c^* = \left( \bm{A}+ i\omega Ra \diag(\bm{\hat{c}}_{\bm{b}}) \sin\theta - Ra {\bm{B}^{-1}}\bm{H}v^* \diag\left(\frac{\bm{\partial\hat{c}_b}}{\bm{\partial\hat{z}}}\right)\right)c^*.
\end{equation}
$c^*$  and $v^*$ are the concentration and velocity eigenfunctions respectively, $\bm{A} = \left(\displaystyle{\frac{\partial^2}{\partial \hat{z}\:^2}} - \omega^2 \times \bm{I} \right)$ and $\bm{B} = \left(\displaystyle{\frac{\partial^2}{\partial \hat{z}\:^2}} - \omega^2 \times \bm{I} \right)$, $\bm{H} = \left( - \omega^2 cos \theta + i\omega sin \theta \displaystyle{\frac{\partial}{\partial \hat{z}}} \right)$ and $\bm{I}$ is an identity matrix. The elements in the top and bottom rows of the coefficient matrices $\bm{A}$ and $\bm{H}$ which represents the Dirichlet boundary condition on concentration are replaced by values for Neumann condition for velocity in $\bm{B}$. Due to the gravity component parallel to the inclined system, it is expected for the growth rate to be complex $(\sigma = \sigma_r+i\sigma_i)$ which modifies the perturbations to take the form $(\hat{v}\:{'}, \hat{c}\:{'})(\hat{x},\hat{z},\hat{t})= (v^*,c^*)(\hat{z},\hat{t}_o) e^{(i(\omega \hat{x}+ \sigma_i \hat{t}) + \sigma_r \hat{t})}$. The imaginary part $(\sigma_i)$ appears as an angular frequency of a progressive wave of wavenumber $\omega$, indicating an $x$-directional motion of the evolving perturbations with a wave velocity, $\hat{v}_w = \sigma_i/\omega$. By choosing $\theta$ = 0, equation (14) becomes the eigenvalue problem $\sigma \bm{c^*} =  \left( \bm{A} - (-\omega^2) Ra \bm{B}^{-1} \diag\left(\displaystyle{\frac{\bm{\partial\hat{c}_b}}{\bm{\partial\hat{z}}}}\right)\right)\bm{c^*}$ derived in previous studies \cite{Lutheretal.2021,RaadHassanzadeh2015} for horizontal systems, where the growth rate is real $(\sigma_i = 0)$.

\begin{figure}[H]
  \centering
  \begin{tabular}[b]{c}
    \includegraphics[width=.45\linewidth]{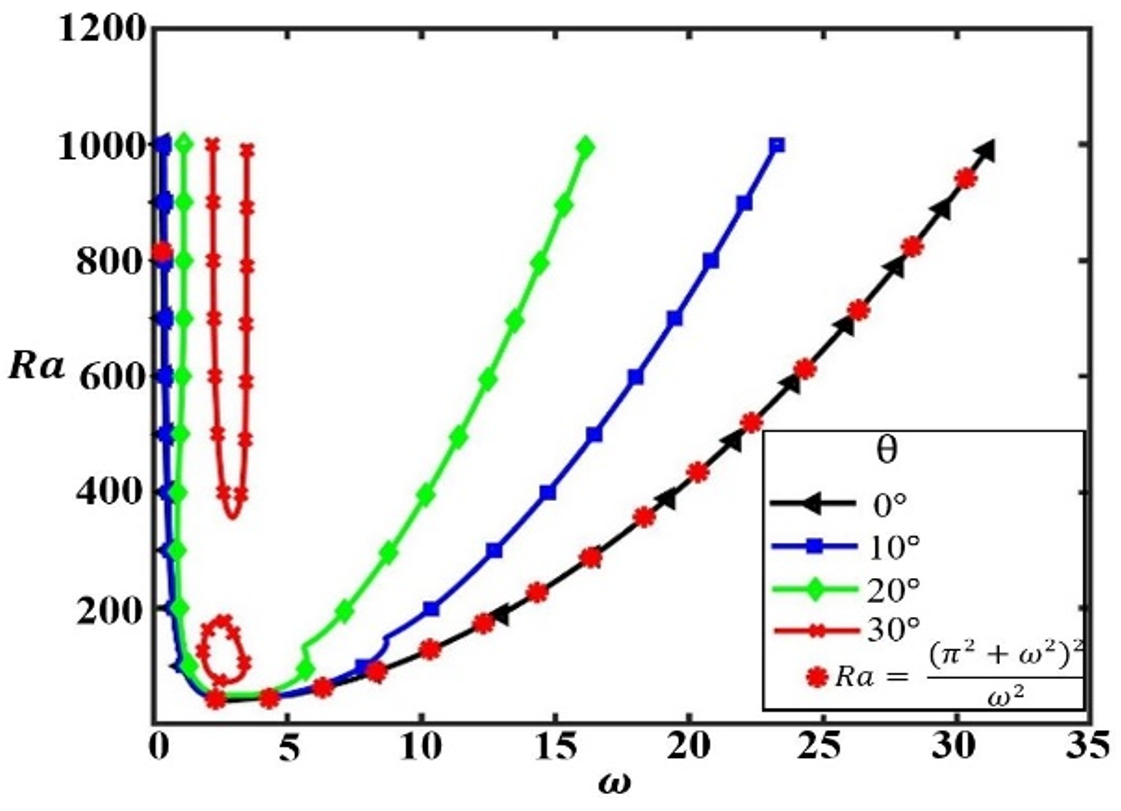} \\
    \small (a)
  \end{tabular} \qquad
  \begin{tabular}[b]{c}
    \includegraphics[width=.45\linewidth]{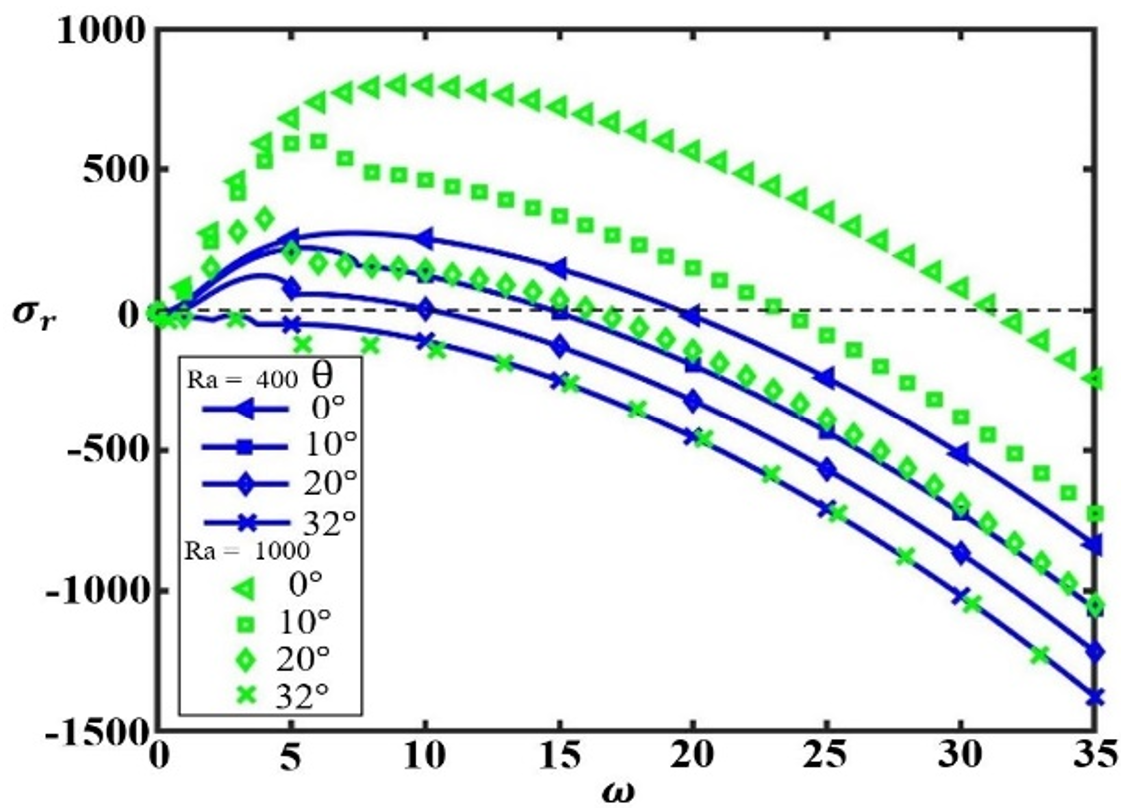} \\
    \small (b)
  \end{tabular}
        \caption{\label{fig:inclined2} (a) Neutral curves for the first mode (n = 1) of the steady convective instability for $0^{\circ}\leq \theta \leq 30 ^{\circ}$. The declining range of the disturbance wavenumber ($\omega$) as $\theta$ approaches the cut-off angle of $31.49^{\circ}$ and the comparison of our result with the classical solution of $Ra = \frac{(\pi^2+\omega^2)^2}{\omega^2}$ for the horizontal case validate our computations. (b) The relationship between the real part of the growth rate (perturbation growth rate) $\sigma_r$ and the wavenumber $\omega$ for various angles at the onset of convective instability ($\hat{t}_o$) for $\theta = 32 ^{\circ}$, indicating instability beyond the cut-off angle for the steady case.}
\end{figure}

To test our formulation, we compare it with the published solution $Ra = \displaystyle{\frac{(n^2 \pi^2+\omega^2)^2}{\omega^2}}$  for the most unstable mode $(n = 1)$ in a horizontal configuration where we specify $\displaystyle{\frac{d\hat{c}_b}{d\hat{z}}}=-1$ for a constant base state in (14) and seek the critical wavenumber $\omega_c$ such that $\sigma=0$ for various $Ra$  (Fig.~\ref{fig:inclined2}a) \cite{HortonandRogers1945,Lapwood1948}. For $\theta = 0$, the minimum of the function $Ra(\omega)$ occurs at $Ra_c = 4\pi^2$ when $\omega_c =\pi$ as in Horton and Rogers\cite{HortonandRogers1945} and Lapwood\cite{Lapwood1948} . For $\theta > 0^{\circ}$, our results are in perfect agreement with the literature for the constant base state problem derived by substituting $\hat{c}_b = 1-\hat{z}$ for the base state and its $\hat{z}$ derivative in (14), as in Rees and Bassom \cite{ReesBassom2000}. Increasing the angle of inclination decreases the growth rate $(\sigma_r)$ of the perturbations, weakening the force that triggers the density driven convection. When $\theta$ increases beyond the critical value of $\theta = 31.49^{\circ}$, the system is stable to two-dimensional perturbations for all values of $Ra$ (Fig.~\ref{fig:inclined2}b) in agreement with Rees and Bassom \cite{ReesBassom2000}. 

\section{\label{sec:level1}results and discussion} 

\begin{figure}[H]
  \centering
  \begin{tabular}[b]{c}
    \includegraphics[width=.45\linewidth]{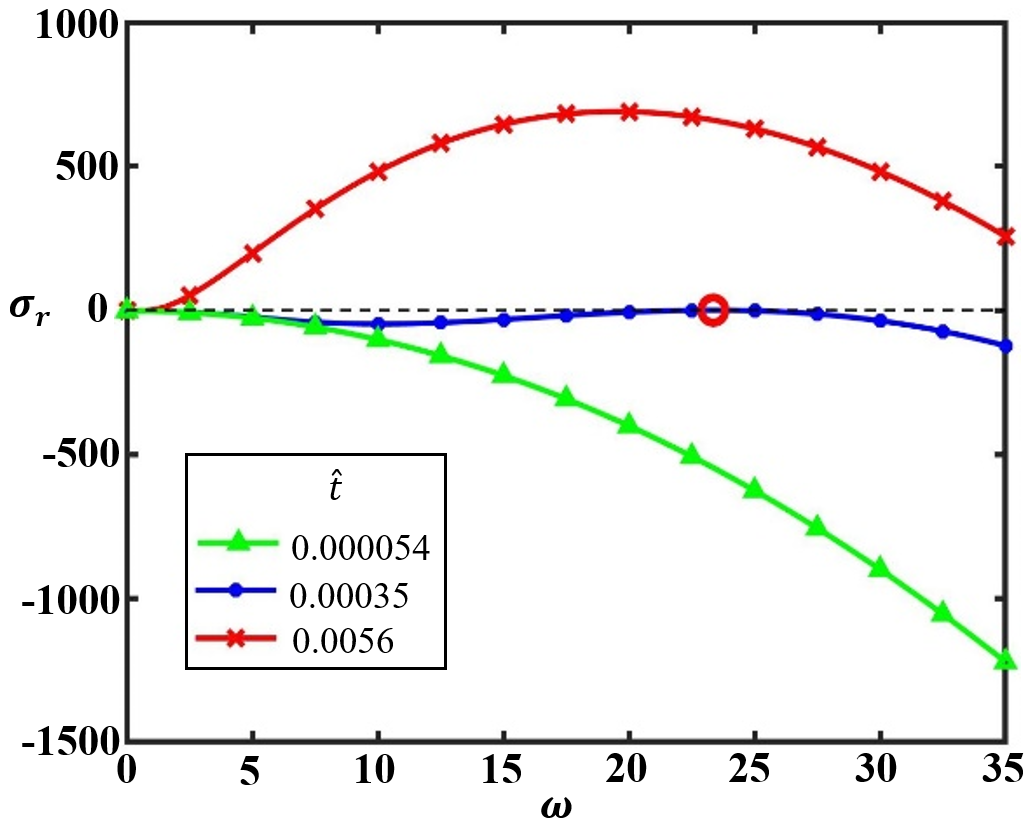} \\
    \small (a)
  \end{tabular} \qquad
  \begin{tabular}[b]{c}
    \includegraphics[width=.45\linewidth]{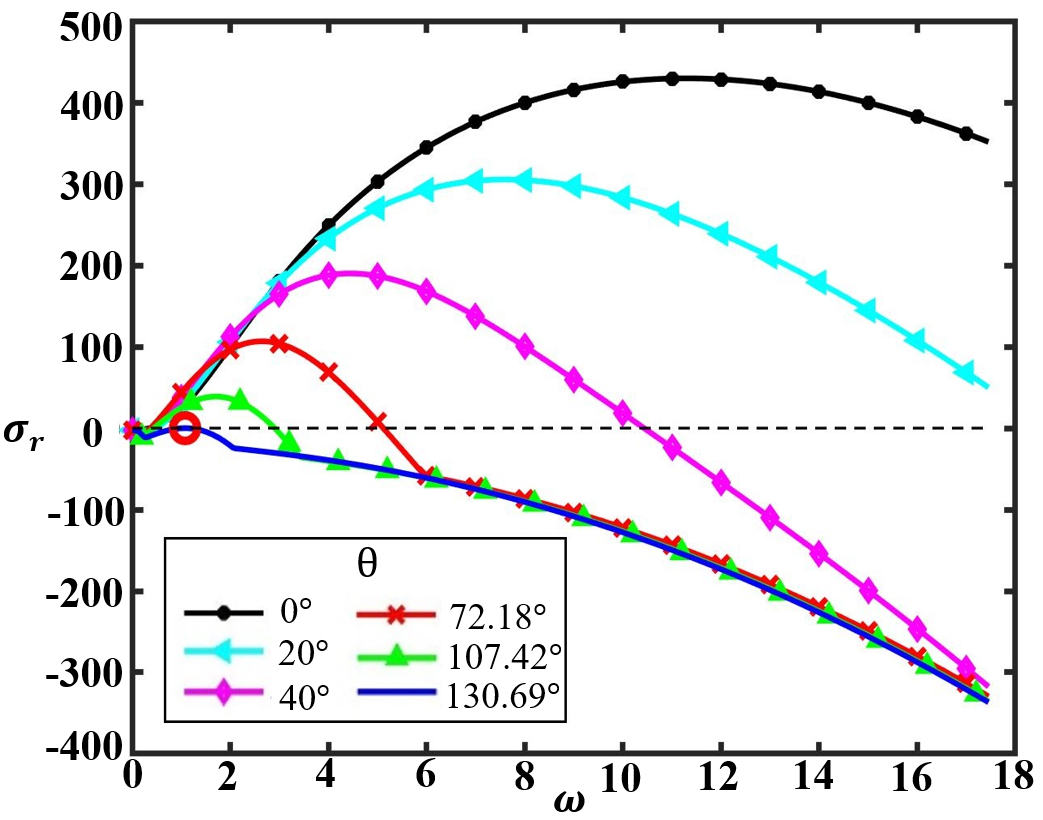} \\
    \small (b)
  \end{tabular}
   \caption{\label{fig:inclined3} (a) The perturbation growth rate ($\sigma_r$) increases in time for the specified wavenumbers ($\omega$) for a horizontal aquifer with $Ra = 400$ and an unsteady base state. The onset time ($\hat{t}_o$) and the critical wavenumber ($\hat{\omega}_o$) are shown in red. (b) Perturbation growth rate ($\sigma_r$) versus wavenumber ($\omega$) for several inclined systems $\theta = 0^\circ$, $20^\circ$, $40^\circ$, $71.18^\circ$, $107.42^\circ$ and $130.69^\circ$ plotted at the onset time ($\hat{t}_o$) for $\theta = 130.69^\circ$, the maximum or cut-off angle for $Ra = 400$ beyond which the system does not become unstable.}
\end{figure}

\begin{figure}[H]
  \centering
  \begin{tabular}[b]{c}
    \includegraphics[width=.45\linewidth]{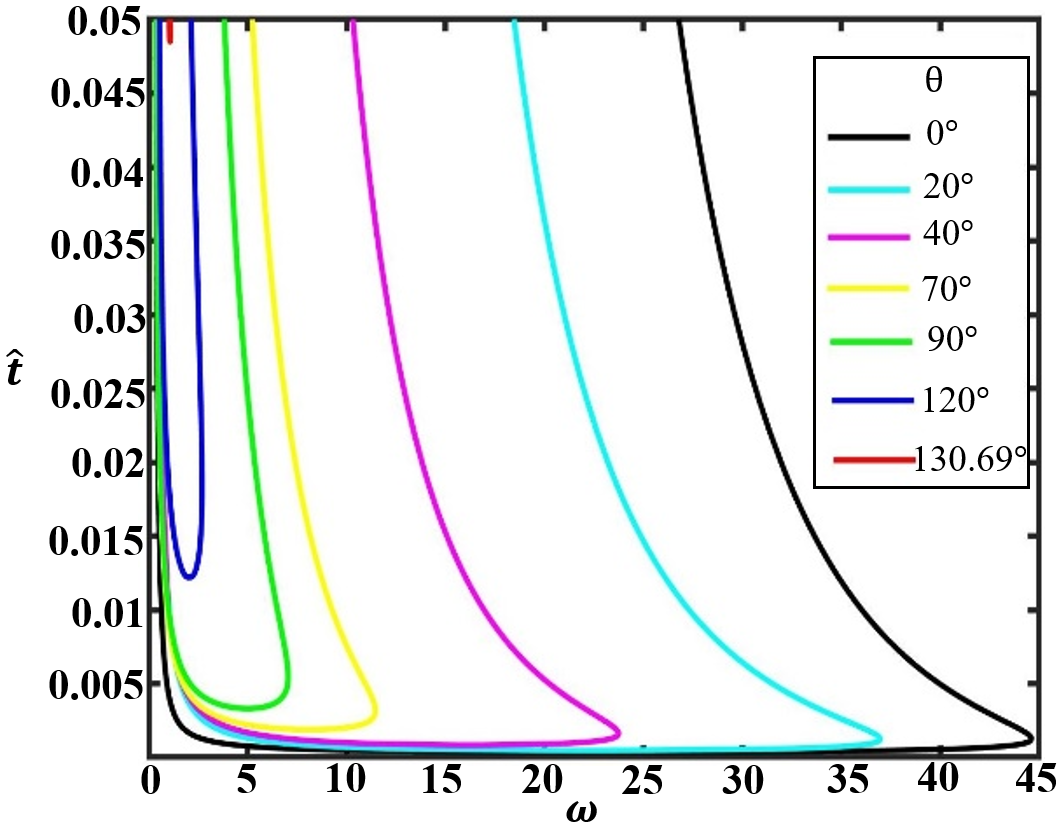} \\
    \small (a)
  \end{tabular} \qquad
  \begin{tabular}[b]{c}
    \includegraphics[width=.45\linewidth]{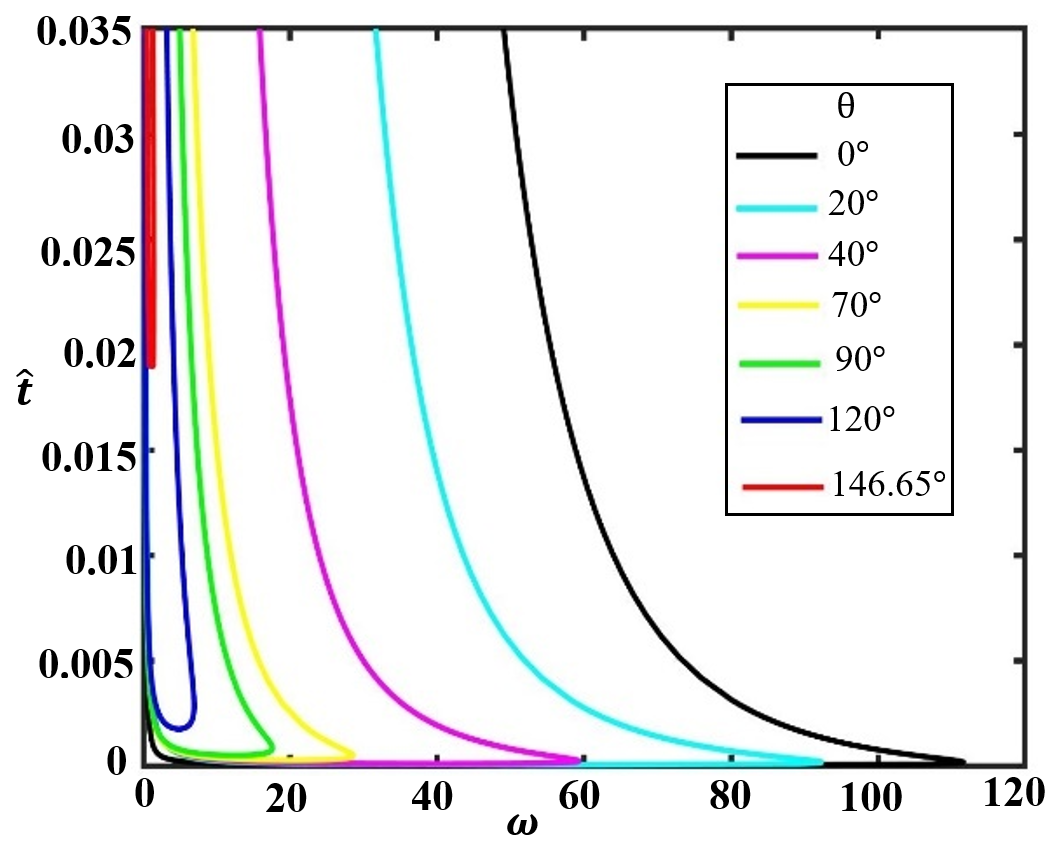} \\
    \small (b)
  \end{tabular}
  \caption{\label{fig:inclined4}
  The time of instability decomposed into Fourier wavenumbers $\omega$ for various angles of inclination ($\theta$) for (a) $Ra = 400$ and (b) $Ra = 1000$, highlighting their cut-off angles in red. The range of the wavenumber, measuring the instability region, declines while the onset time ($\hat{t}_o$) increases with $\theta$, delaying the occurrence of instability.}
\end{figure}

The steady and transient cases are compared in Figs.~\ref{fig:inclined2} -~\ref{fig:inclined3}. The growth rate of perturbations in both cases is amplified by increasing $Ra$ and inhibited by increasing the angle of inclination. LSA predicts a range of wavenumber that grows; the fastest growing wavenumber has the largest growth rate. Expectedly, the neutral stability curves (Fig.~\ref{fig:inclined2}a) and the growth rate profile $\sigma_r(\omega)$ (Fig.~\ref{fig:inclined2}b) for the steady base state do not depend on time in contrast with the transient case where the growth rate of the perturbations does evolve in time (Fig.~\ref{fig:inclined3}a). 
The range of disturbance wavenumber observed for horizontal media ($\theta = 0^{\circ}$) in Fig.~\ref{fig:inclined4} results from the suppression of short waves by diffusion and the inability of a relatively thin diffusive boundary layer to propagate long waves \cite{Slim2014}. As the angle of inclination increases, this range gets smaller (Fig.~\ref{fig:inclined4}).

\begin{figure}[H]
  \centering
  \begin{tabular}[b]{c}
    \includegraphics[width=.45\linewidth]{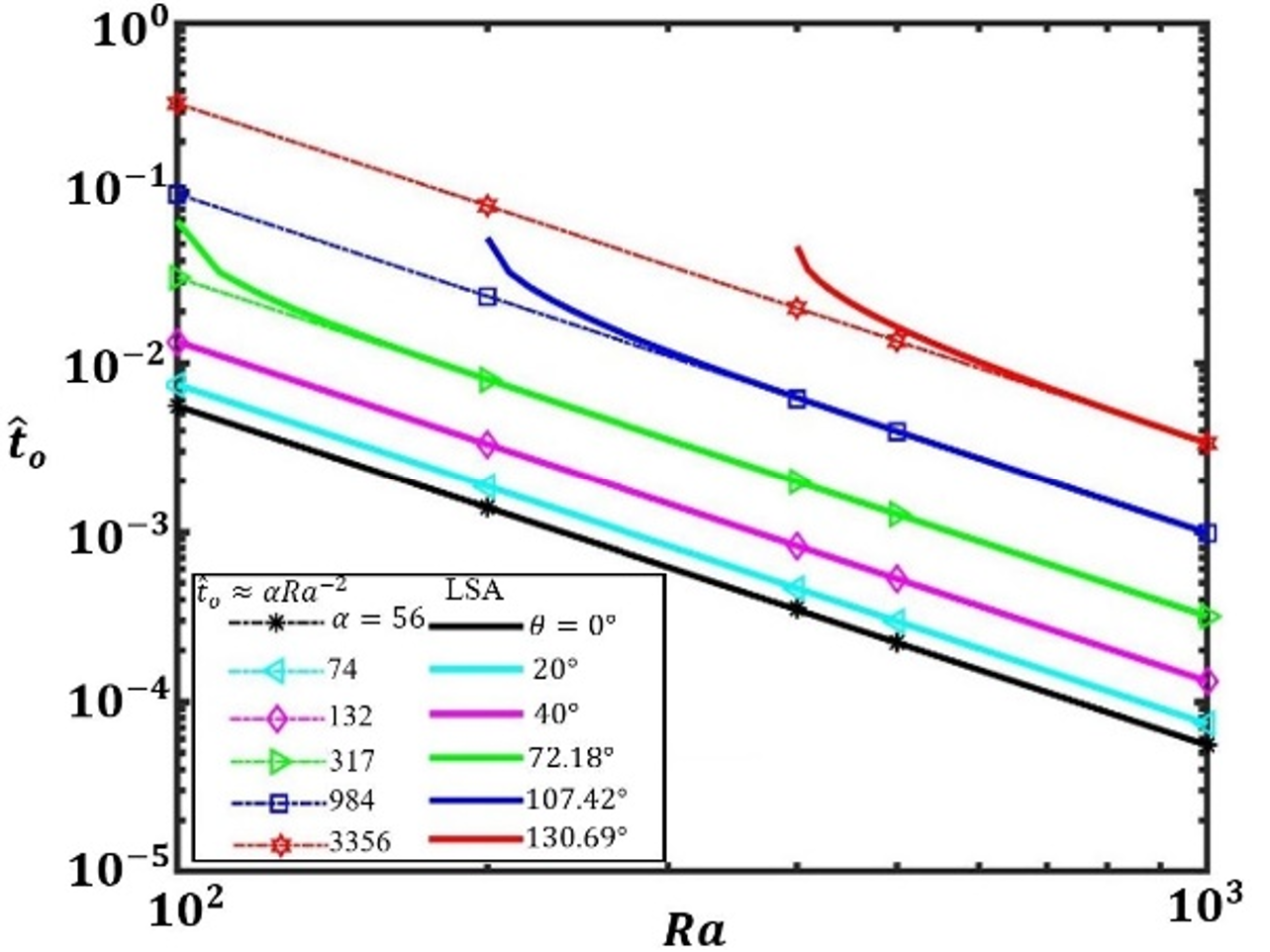} \\
    \small (a)
  \end{tabular} \qquad
  \begin{tabular}[b]{c}
    \includegraphics[width=.45\linewidth]{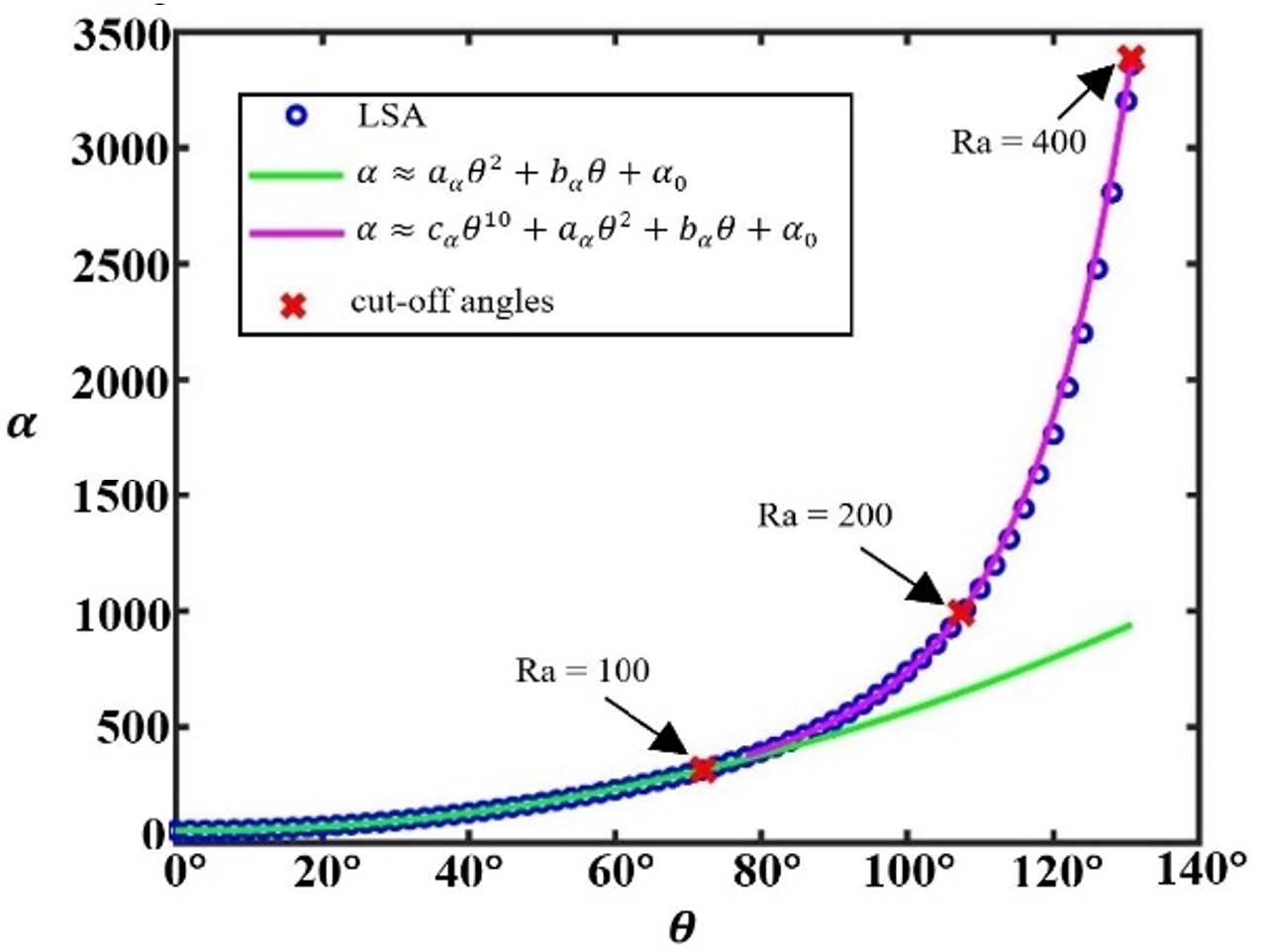} \\
    \small (b)
  \end{tabular}
\caption{\label{fig:inclined5} (a) The relationship between the onset time ($\hat{t}_o$)and Rayleigh number for $\theta = 0^\circ - 130.69^\circ$. By having a larger onset time near the cut-off angles, this relationship deviates from $\hat{t}_o  = \alpha Ra ^ {-2}$, the scaling for large $Ra$ limits. (b) The prefactor $\alpha(\theta)$ for the onset time increases $\theta$ for large Rayleigh limit beyond the region of influence of the cut-off angle. $\hat{t}_o$ increases with $\theta$. $a_\alpha =0.054$ , $b_\alpha = -0.2$ , and $c_\alpha =1.68 \times 10^{-18}$. The cut-off angles for $Ra = 100$, $200$, and $400$ are marked with the red crosses.}
\end{figure}

For horizontal media, the growth rate is real and negative during early time when the perturbations decay but becomes positive at the critical wavenumber as the onset time is crossed. In this case, the relationship obtained between the onset time of instability ($\hat{t}_o$) and Rayleigh number, $\hat{t}_o = \alpha_0Ra^{-2}$, is in agreement with previous studies (Fig.~\ref{fig:inclined5}a). Our constant of proportionality $(\alpha_0\approx56)$ agrees with results in other studies that employ QSSA in normal coordinates but can take a different value depending on the assumptions made \cite{TiltonDanielRiaz2013}. 

The scaling relationship $\hat{t}_o  = \alpha Ra ^ {-2}$ holds for $\theta> 0^{\circ}$, except when $\theta$ approaches the cut-off angle (Fig.~\ref{fig:inclined5}). Below the cut-off angle, the gravitational force that drives density driven convection declines as $\theta$ increases, delaying the onset of instability. Beyond the cut-off angle (which depends on $Ra$), no instability arises. The prefactor increases with the angle of inclination and a reasonably accurate polynomial fit to our numerically computed results are $\alpha(\theta) \approx 0.054 \theta^{2} - 0.28 \theta + \alpha_o$ for $ 0^{\circ} \leq \theta \leq 76^{\circ}$, with an additional higher order term 
$\alpha(\theta) \approx 1.68 \times 10^{-18} \theta^{10} + 0.054 \theta^{2} - 0.28 \theta + \alpha_o$ for $ 76^{\circ} \leq \theta \leq 230.69^{\circ}$ (Fig.~\ref{fig:inclined5}b), where $\alpha_0\approx56$ is the horizontal prefactor. This behavior differs from that in porous systems with steady base state, where instability is not observed for $\theta > 31.49^{\circ}$ regardless of $Ra$ (see Rees and Bassom \cite{ReesBassom2000}).

\begin{figure}[H]
  \centering
  \begin{tabular}[b]{c}
    \includegraphics[width=.45\linewidth]{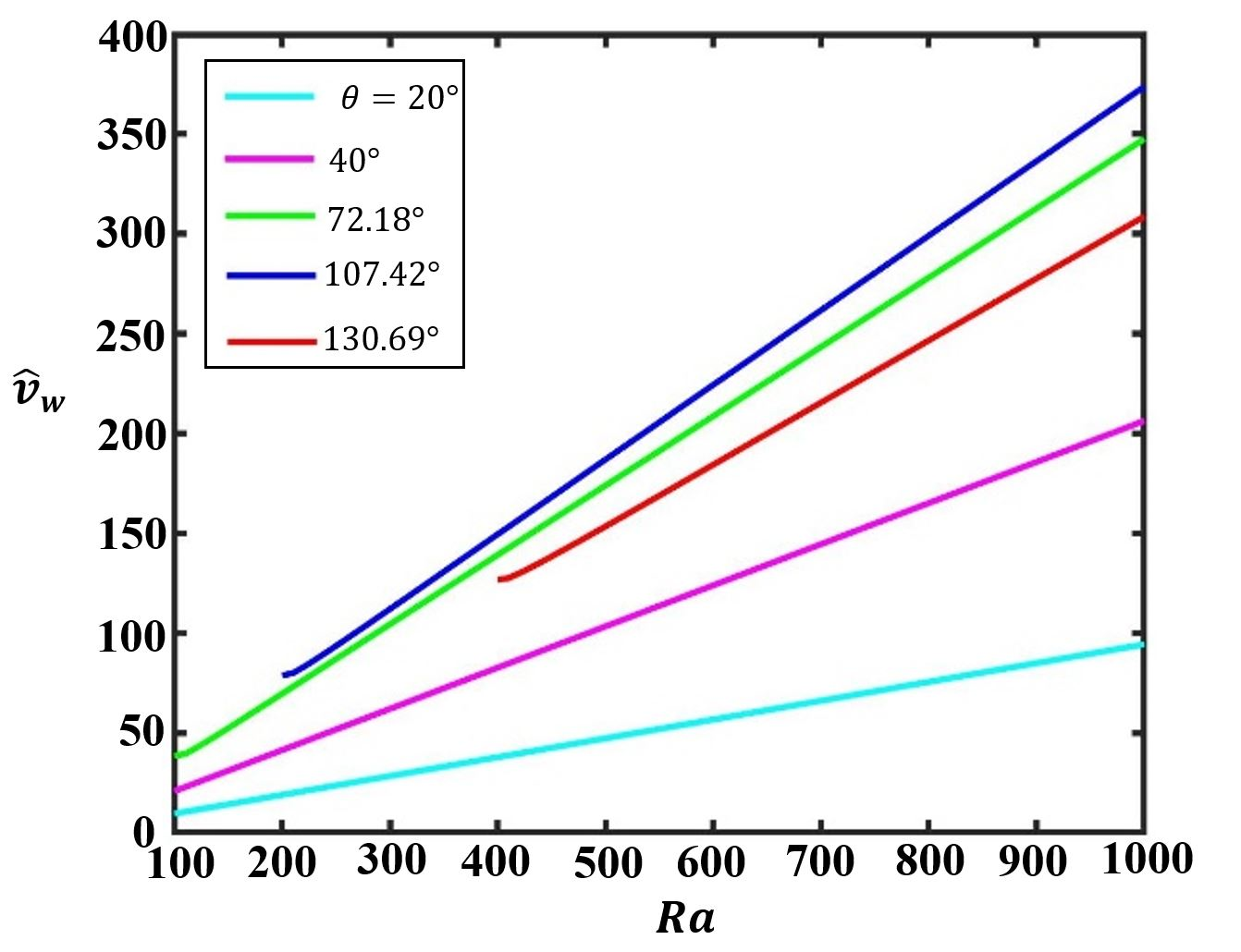} \\
    \small (a)
  \end{tabular} \qquad
  \begin{tabular}[b]{c}
    \includegraphics[width=.45\linewidth]{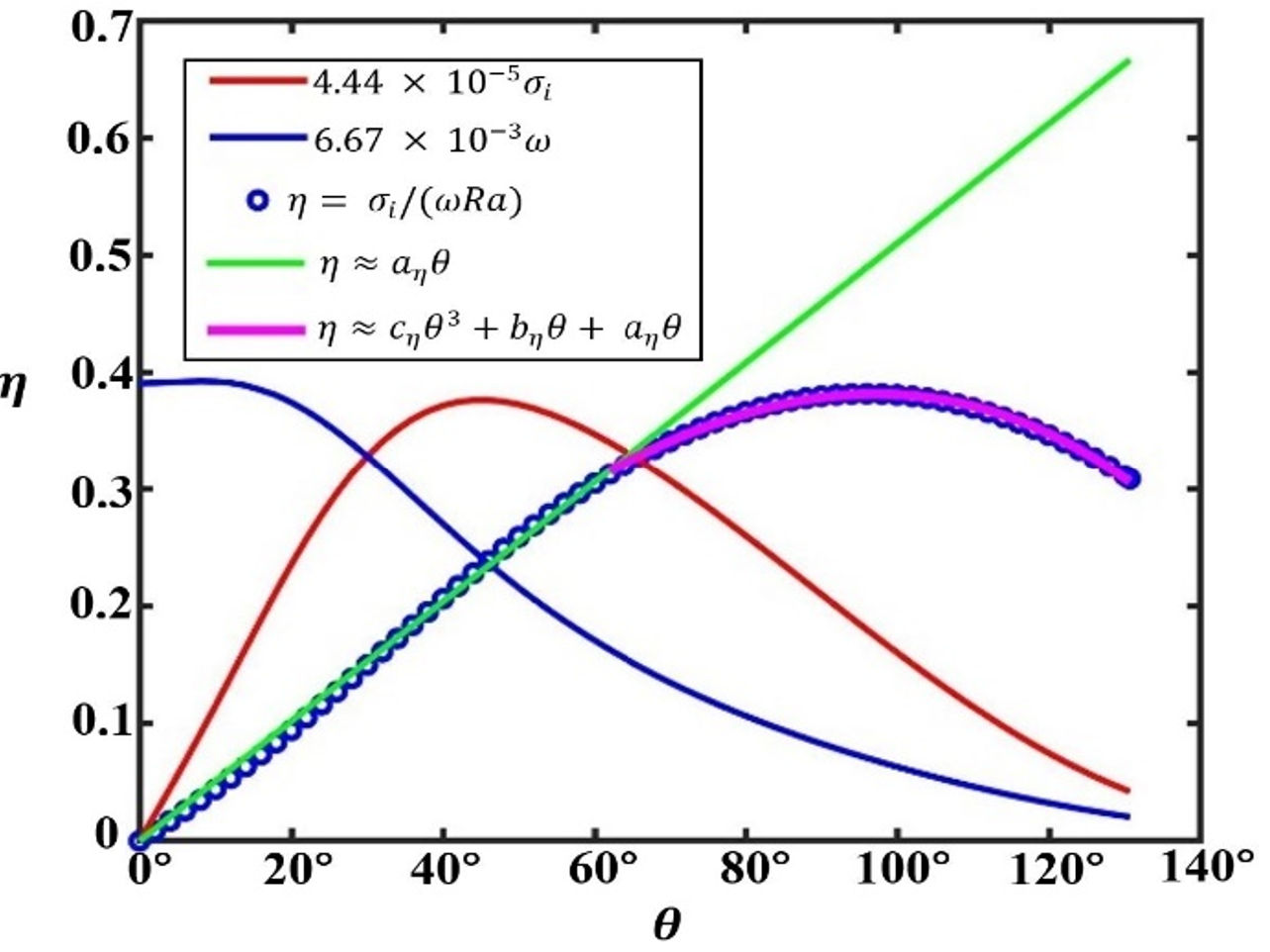} \\
    \small (b)
  \end{tabular}
\caption{\label{fig:inclined6} (a) The lateral wave velocity ($\hat{v}_w$) of the instabilities for various angles $\theta$ scales linearly with Rayleigh number $Ra$ as $\hat{v}_w = \eta Ra$, where $\hat{v}_w = \sigma_i/\omega$, except at the cut-off angle of the $Ra$. (b) The prefactor $\eta(\theta)$ of $\hat{v}_w$ is a non-monotonic function of $\theta$, becoming maximum at $96^{\circ}$, where $a_\eta = 0.0051$, $b_\eta = 0.00079$, and $c_\eta = -2.08 \times 10^{-7}$ are the coefficients of the polynomial fits in the plot. The wavenumber and the imaginary part of the growth rate $\sigma_i$ are plotted in blue and red respectively. $\sigma_i$ becomes maximum at $45^{\circ}$ where the gravity components are equal.}
\end{figure}

The growing instabilities at the onset time are not stationary due to the gravity component parallel to the layer, unlike in the horizontal case where gravity is perpendicular making the perturbations non-oscillatory $(\sigma_i=0)$. The instabilities travel laterally with velocity $\hat{v}_w$ which is proportional to $Ra$ and $\theta$ as $\hat{v}_w \approx \eta (\theta) Ra$, where $\eta \approx 0.0051 \theta$ for $0^{\circ} \leq \theta \leq 58 ^{\circ}$ and $ \eta \approx -2.08 \times 10^{-7} \theta^3  + 0.00589\theta$ for $58 ^{\circ} < \theta \leq 130.69^{\circ}$ (Fig.~\ref{fig:inclined6}). The velocity $\hat{v}_w$ increases to a maximum at $96^{\circ}$, which is close to but not exactly at $90^{\circ}$ where the vertical gravity component is maximum. The reason for the difference is that the onset time is also dependent on $\theta$, becoming later as $\theta$ increases. Beyond this maximum value, the velocity declines (Fig.~\ref{fig:inclined6}). The magnitude of $\sigma_i$ is maximum at $\theta = 45^{\circ}$ where the magnitude of the two gravity components are equal to each other.

For sufficiently large $Ra$ (possibly exceeding the expected range for CO$_2$ sequestration), the system can be unstable for $\theta=179^{\circ}$ (see Appendix) due to $\sigma_i \neq 0$, even though the configuration is gravitationally stable.  However, when $\theta=180^{\circ}$ the system remains stable regardless of $Ra$. 

For moderate $Ra$ $(100-1000)$, a generalized form which includes the effect of inclination can be formulated for the dimensional onset time and wave velocity as  $t_o \approx \alpha(\theta)(\phi \mu \sqrt{D}/k \Delta \rho g)^2$ and $v_w \approx \eta (\theta)k \Delta \rho g / \mu$, respectively, where $\Delta \rho = \rho_o\beta c_s $. The instability found for $\theta \geq 90^{\circ}$ where the fluid configuration is gravitationally stable is purely due to the base velocity shear, since the base flow is faster near the source where the concentration is maximum $(\hat{u}_b \sim \hat{c}_b$), generating a Kelvin-Helmholtz type instability.

\section{\label{sec:level1}Conclusion}

This study exposes the role of porous medium inclinations on the onset of density driven convective instability using linear stability analysis. The quasi-steady state approximation in normal coordinates was introduced to handle the time-dependency of the diffusing base state, which is also non-stationary. Our analysis indicates that the onset time is delayed as the inclination increases, until it reaches a Rayleigh-dependent cut-off angle beyond which the system remains stable. At the onset time, the growing perturbations migrate laterally with a velocity that depends non-monotonically on the angle of inclination and linearly on the Rayleigh number. For gravitationally stable configurations including angles between $90^{\circ}$ and $179^{\circ}$, perturbations grow due to the velocity shear produced from the non-uniform base velocity characterizing Kelvin-Helmholtz type instability. Further exploration of absolute instabilities and non-linear effects not captured by our analysis remains an interesting subject for future work.

\begin{acknowledgments}
The authors gratefully acknowledge the support of Petroleum Technology Development Fund (PTDF) in Nigeria. 
\end{acknowledgments}

\appendix*
\section{Stability for \bm{$\theta = 179^{\circ}$}}

\begin{figure}[H]
  \centering
  \begin{tabular}[b]{c}
    \includegraphics[width=.45\linewidth]{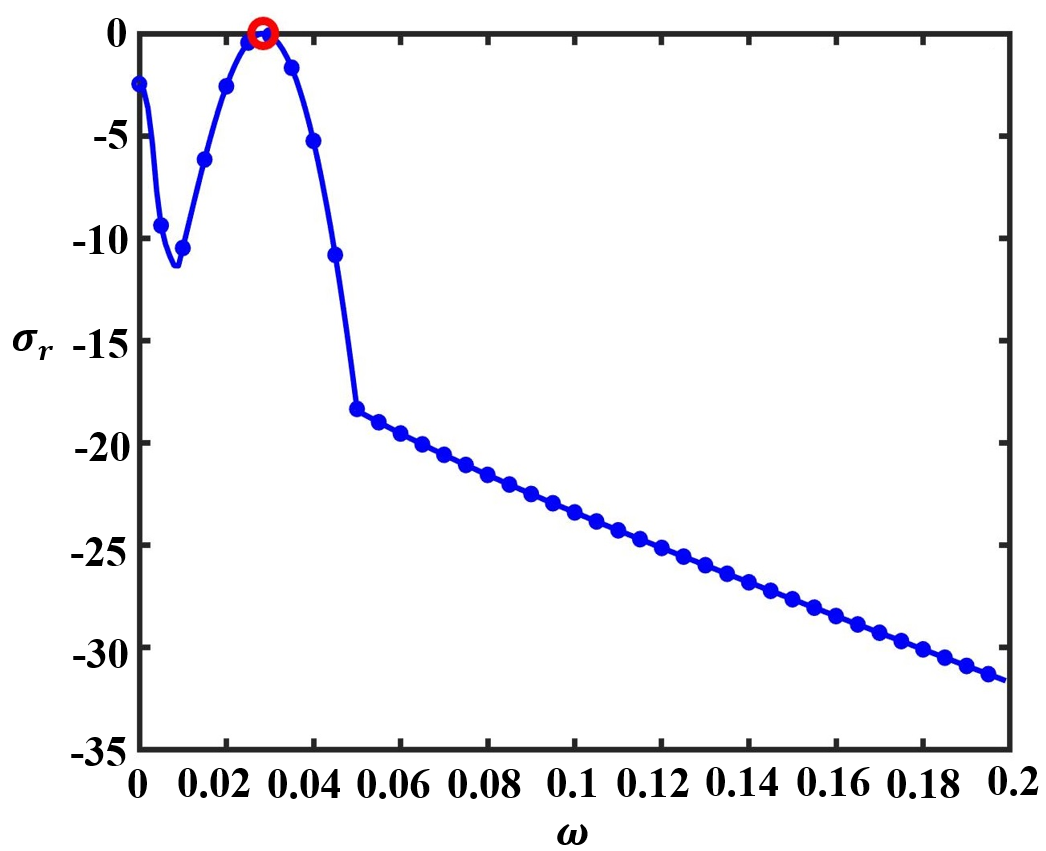} \\
    \small (a)
  \end{tabular} \qquad
  \begin{tabular}[b]{c}
    \includegraphics[width=.45\linewidth]{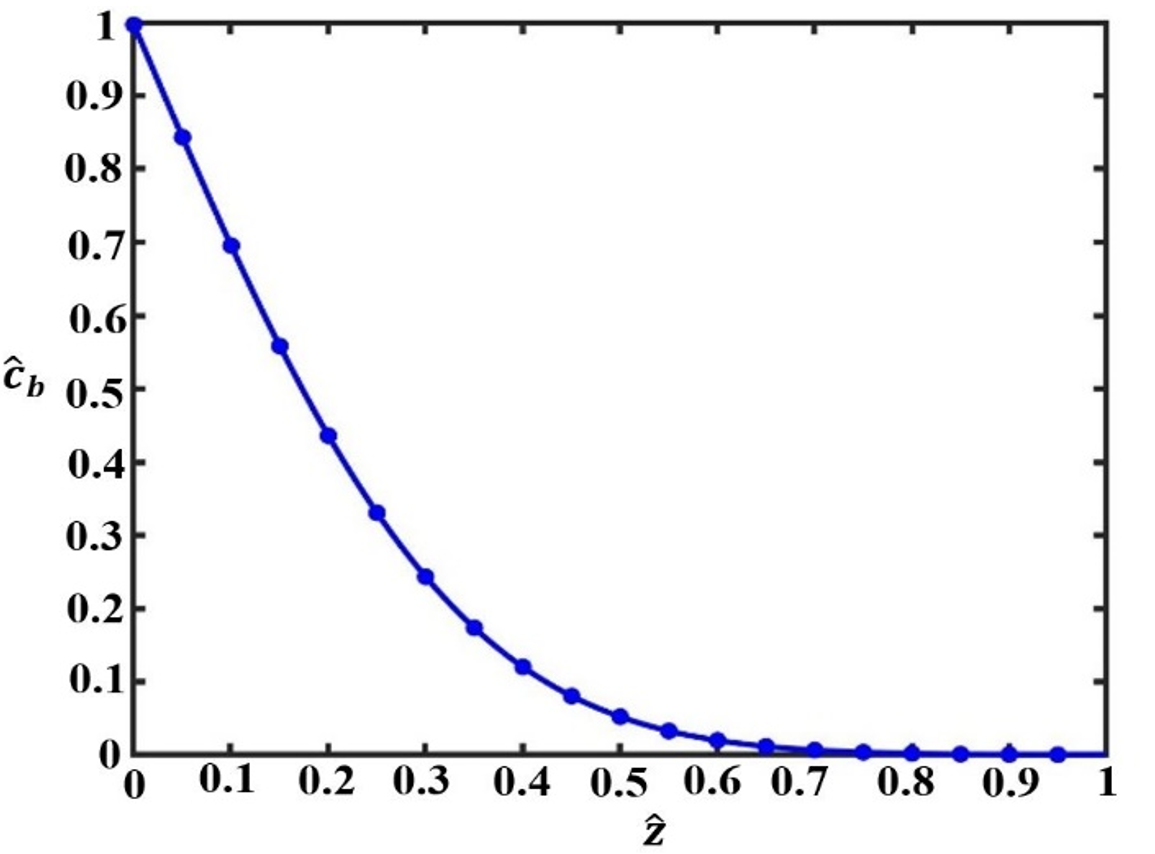} \\
    \small (b)
  \end{tabular}
 \caption{\label{fig:inclined7} (a) 
  Perturbation growth rate ($\sigma_r$) versus wavenumber ($\omega$) showing the critical conditions in red for $\theta = 179^\circ$ and $ Ra = 9 \times 10^5$, indicating the existence of instabilities. This suggest the lack of a global cut-off angle, implying that a diffusive boundary layer will become unstable when $Ra$ is sufficiently large. (b) The diffusive profile at the onset time. Both plots are made at $\hat{t}_o = 0.033$.}
\end{figure}

Here, we show that the system can become unstable for $\theta = 179^{\circ}$ when $Ra = 9 \times 10^5$ by examining the growth rate of different wavenumbers around the onset time; the plot of the corresponding diffusive profile is shown (Fig.~\ref{fig:inclined7}). This indicates the non-existence of a global cut-off angle where the system is stable for any $Ra$, depicting a fundamental difference between transient and steady convective instability in an inclined layer.

\bibliography{PoF_inclinedinstability}

\end{document}